\documentclass[apj]{emulateapj}
\usepackage{graphicx,latexsym,natbib}

\begin{document}

\title{\textit{Swift} XRT Timing Observations of the Black-Hole Binary SWIFT J1753.5--0127:
Disk-Diluted Fluctuations in the Outburst-Peak}
\author{M. Kalamkar\altaffilmark{1}, M. van der Klis\altaffilmark{1}, P. Uttley\altaffilmark{1}, Diego Altamirano\altaffilmark{1}, Rudy Wijnands\altaffilmark{1}}

\altaffiltext{1}{Astronomical Institute, ``Anton Pannekoek'', University of Amsterdam, Science Park 904, 1098 XH, Amsterdam, The Netherlands}

\email{m.n.kalamkar@uva.nl}

\begin{abstract}
\noindent After a careful analysis of the instrumental effects on the Poisson noise to demonstrate the feasibility of detailed stochastic variability studies with the \textit{Swift} X-Ray Telescope (XRT), we analyze the variability of the black hole X-ray binary SWIFT J1753.5-0127 in all XRT observations during 2005-2010. We present the evolution of the power spectral components along the outburst in two energy bands: soft (0.5--2 keV) and hard (2--10 keV), and in the hard band find results consistent with those from the \textit{Rossi X-ray Timing Explorer} (RXTE). The advantage of the XRT is that we can also explore the soft band not covered by RXTE. The source has previously been suggested to host an accretion disk extending down to close to the black hole in the low hard state, and to show low frequency variability in the soft band intrinsic to this disk. Our results are consistent with this, with at low intensities stronger low-frequency variability in the soft than in the hard band. From our analysis we are able to present the first measurements of the soft band variability in the peak of the outburst. We find the soft band to be less variable than the hard band, especially at high frequencies, opposite to what is seen at low intensity. Both results can be explained within the framework of a simple two emission-region model where the hot flow is more variable in the peak of the outburst and the disk is more variable at low intensities.\\
\end{abstract}


\section{Introduction}\label{intro}

\noindent Transient stellar-mass black hole candidate X-ray binaries (BHBs) in outburst broadly exhibit two different states: low luminosity hard states in which the spectrum is dominated by hard power law emission (out to a few tens of keV) from a hot inner flow, inner corona, and/or the base of a jet, and high luminosity soft states in which the spectrum is dominated by black-body emission (peaking at up to a few keV) from the accretion disk, often accompanied by a power law tail.  There is no agreement about the exact structure and origin of the power law emitting region, nor on the disk geometry: while previously it was thought that in the hard state the disk is truncated at a large distance of at least several tens of gravitational radii, recently it has been debated if instead it extends all the way down to the innermost stable circular orbit (ISCO). See \citet{belloni2005}, \citet{klis2006} and \citet{remillard2006} for detailed descriptions of the different states along an outburst and \citet{done2007} for a detailed discussion of the different models. \\
\\ 
The  ubiquitous X-ray variability in BHBs could provide useful constraints on the structure and behavior of the emitting regions, beyond that provided by the X-ray spectrum.  \citet{lyub1997} proposed a model which is becoming widely accepted (e.g. \citealt{chu2001,uttley2005,done2007}), where the variability originates as mass-accretion fluctuations in the accretion flow, which propagate towards and through the X-ray emitting regions to produce the observed flux variability.  Since the variability in the hard states is strong (few tens of percent fractional rms) while in the soft states it is much weaker (a few percent),  \citet{chu2001} suggested that the hot, power law emitting inner flow is responsible for this variability, while the cool disk is stable.   Furthermore, they showed that in a `standard' geometrically-thin cool disk, high frequency fluctuations can survive only if they arise at small radii, as at larger radii, viscous damping will prevent them from propagating inward. In the hard state, the power spectrum consists of broad band-limited noise accompanied by Quasi Periodic Oscillations (QPOs). The low frequency break in the noise power spectrum could be associated with the transition radius to the hot flow at the inner edge of the cool disk, the QPO could be the Lense-Thirring precession of the hot flow, and the high frequency break may be associated with the innermost radius of the hot flow (\citeauthor{stella1998} 1998, \citeauthor{chu2001} 2001).\\
\\
The \textit{Rossi X-ray Timing Explorer} (RXTE) Proportional Counter Array (PCA) was sensitive in the $>$ 2 keV band dominated by the power law emission. Most variability studies until recently were done with RXTE, so the evolution of the variability below 2 keV remained unexplored. Recent work by \cite{wilkinson} and \cite{uttley2011} using \textit{XMM-Newton} showed that in the hard state of the BHBs SWIFT J1753.5-0127, GX~339-4 and Cyg~X-1 the disk contributes significantly to variability in the soft band ($<$ 2 keV) at Fourier frequencies below 1~Hz. Moreover, the disk variations precede the correlated power law variations by a few tenths of a second. These results cast doubt on the notion of a stable disk not contributing to the source variability, implying that (at least in the hard state and at low frequencies) substantial flux variability is contributed by the cool disk. This highlights the importance of exploring the variability in the soft band.  Like \textit{XMM-Newton}, the \textit{Swift} X-ray Telescope (XRT) has a lower energy bound (0.3 keV) than the RXTE PCA, providing another opportunity to explore the variability at lower energies dominated by the disk emission. This directly addresses the question of the origin of the variability. \\

\begin{figure}
\includegraphics[width=6.3cm,height=8.8cm,angle = -90]{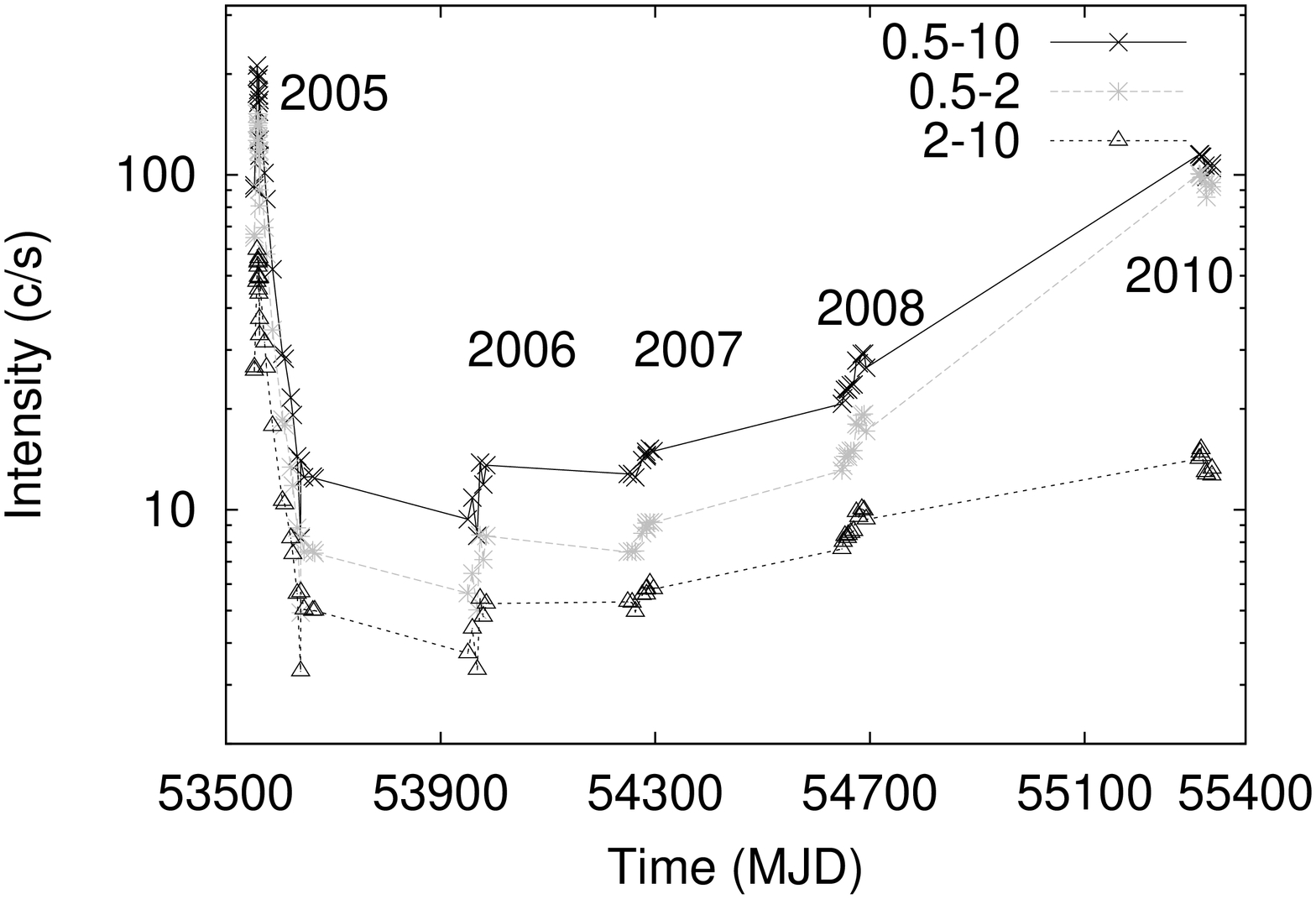}
\includegraphics[width=6.05cm,angle = -90]{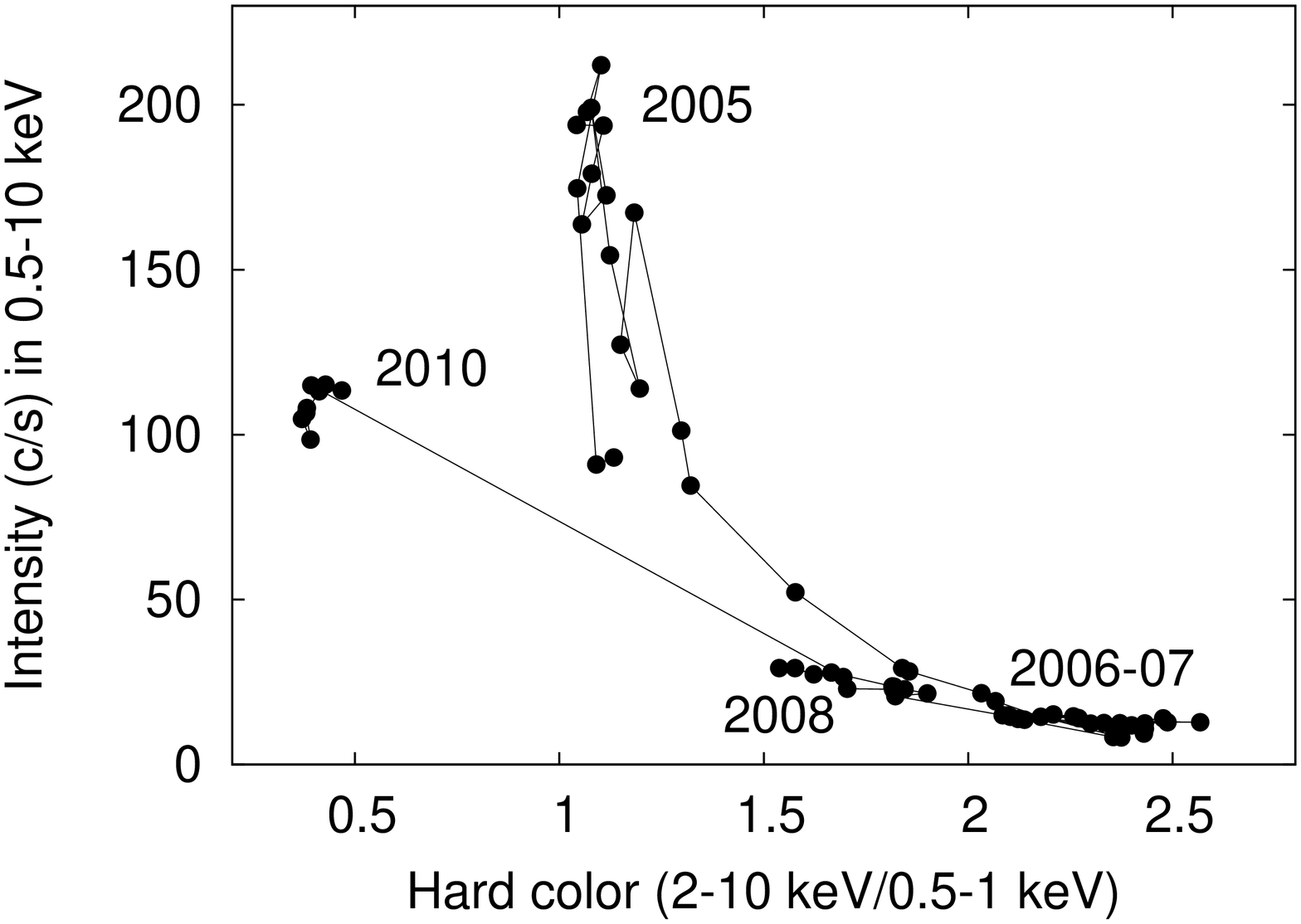}
\caption{Light curve of the outburst during 2005-2010 (top) and corresponding hardness-intensity diagram (bottom) in the energy bands indicated. Each point represents one observation. }
\label{lchc}
\end{figure}
\noindent SWIFT J1753.5-0127 was discovered with the \textit{Swift} Burst Alert Telescope on 2005 May 30 \citep{palmer2005} and has been in outburst since then. The source remained in the hard state (\citeauthor{miller2006} 2006 and \citeauthor{cadolle} 2007, \citeauthor{zhang2007} 2007, \citeauthor{ramadevi} 2007, \citeauthor{chiang2010} 2010, henceforth  C07, Z07, RS07 and Ch10, respectively) until June 2009 (MJD~55010) but then was in a softer state until at least July 2010 (MJD~55404) (\citealt{soleri2012}).  During 2005-06, RXTE observations show white noise (frequency independent) at low (0.2--0.4~Hz) frequencies, a QPO ($\sim$0.7 Hz) and red noise (power decreases with frequency) at high (1.5--2~Hz) frequencies, which is typical of the hard state (Z07, RS07). Spectral analyses with \textit{XMM-Newton} and RXTE suggested the presence of a cool accretion disk extending down to near the ISCO that could be modelled with a disk black-body component in addition to the power law (\citeauthor{miller2006} 2006, \citeauthor{reis2010} 2010, Ch10). This is at variance with the picture in which the disk is truncated at larger radii in the hard state. There are also arguments against the presence of a soft disk-like component (see, e.g., \citeauthor{done2007} 2007 and \citeauthor{hiemstra2009} 2009), which, however, do not refute the evidence for intrinsic disk variability found by \citet{wilkinson}.\\
\\
In this work, we use the \textit{Swift} XRT to study the energy-dependent behavior of the variability  of the BHB SWIFT~J1753.5-0127. While some XRT power spectra of BHBs were  previously reported (e.g., \citealt{kennea1659}, \citealt{curran1752-swift}), the present work is the first study of the energy dependence extending below 2~keV of all power spectral components in a BHB along an outburst.  As a prerequisite to this work, we study the instrumental effects on the Poisson-noise spectrum (see Appendix) and demonstrate that detailed variability studies are feasible with \textit{Swift} XRT. In Section \ref{obdata}, we describe the observations and data analysis. In Section \ref{lchd}, we discuss the evolution of the outburst using light curves and the hardness intensity diagram. In Section \ref{timeanalysis}, we first investigate the energy dependence of the variability in a model-independent way, and then fit Lorentzians to the power spectra of all observations to study the evolution of the components and their energy dependence.  We also study the energy dependence of the power spectra inside the 0.5--2 keV band. Finally, in Section \ref{correlations} we study the correlations between the variability parameters and source intensity. In Section \ref{discussion}, we discuss the interpretation of our results, and their model implications. 

\section{Observations and data analysis}\label{obdata}

\noindent We analysed all 66 observations taken in the Windowed Timing (WT) mode  with the X-Ray Telescope \citep[XRT;][]{burrows2005} on board the \textit{Swift} satellite between July 1, 2005 and May 21, 2010. Observations lasted between 0.1 and 11.0 ksec, and contained between 1 and 34  Good Time Intervals (GTIs) of 0.1--2.0 ksec. The CCD was operated at a time resolution of 1.766 ms. We processed the raw data with the task \textit{xrtpipeline} v0.12.6 using standard quality cuts and only grade 0 events. The data were extracted in two ways.  In the first extraction, the  count rates in 0.5--2, 2--10 and 0.5--1 keV energy bands were extracted using the method suggested by \cite{evans2007} for the light curves and hardness ratio. All the intensities reported throughout the paper (and in the figures) are pile-up \footnote{Two or more photons registered as a single higher-energy photon is called pile-up} and bad-column corrected\footnote{Some of the CCD pixels are not used to collect data. Hence, the flux loss caused by this is corrected} and background subtracted, unless otherwise stated.\\
\\
For the second extraction, aimed at generating the power spectra, we first determined the source region on the CCD. As there is no point-spread function for the WT mode, we fitted a Lorentzian to obtain the source centroid. Reduced $ \chi^{2}$ values were 0.5--3 (193 degrees of freedom), occasionally much higher due to bad pixels; in practice this method always sufficed for our purpose. The Lorentzian full width at half maximum  (FWHM) is typically 5 pixels. A 61-pixel (144") source region was selected centered on the centroid pixel, plus, contiguous with this and on the side of the image that is larger, a background region of the same size (or smaller if the edge of the image was reached). Count rates exceeding 150~c/s are at risk of pile-up \citep{evans2007}. To mitigate this, for each GTI the central pixel and, if necessary, additional pairs of pixels symmetrically around it are removed until the count rate is below 150~c/s, basically censoring the piled-up data \citep{evans2007}. This introduces  artificial jumps in count rate between GTIs; the power spectra are not affected by this, as they never straddle GTIs.   \\
\\
Leahy-normalized \citep{leahy1983} fast Fourier-transform power spectra were calculated of 28.93-s continuous intervals,  giving 34.6-mHz as the lowest frequency bin. For the WT mode which has a time resolution of 1.766 ms, the Nyquist frequency is 283.126 Hz.  The power spectra were then averaged per {\it Swift} observation. We also calculated power spectra averaged over multiple observations using longer (57.86-s) continuous intervals. No background correction was applied prior to the generation of the power spectra. To facilitate comparison with RXTE, two energy bands were used: 2--10 keV (also covered by RXTE) and 0.5--2 keV (not covered by RXTE), and occasionally also narrower bands below 2~keV.  \\
\\
The power spectra are affected by two processes: a) pile-up causes the Poisson level to fall below the theoretical value of 2.0 (also reported in \textit{Chandra} data, \citealt{tomsick2004}) and, b) the readout method in the WT mode causes a power drop off at high frequencies (see the Appendix for details). Hence, we analysed the power spectra in the frequency range $<$100~Hz only. The Poisson noise level is estimated by fitting a constant in the 50-100 Hz frequency range, where no source variability is observed and subtracted from the power spectrum. The resulting power spectrum is expressed in source fractional rms normalization. In this normalization, the rms amplitude is expressed as a fraction of the source count rate, instead of the total count rate. To obtain this, each power (after subtracting the Poisson level) is multiplied by a factor $(S+B)/S^2$, where S and B are the source and background count rates, respectively \citep{klis1995}. The fractional rms amplitude is the square root of this 'rms normalized' integral power. It should be noted that the intensities used for renormalization are the ones obtained from the data used for generating the power spectrum, i.e. with the piled-up data removed. The power spectrum of each observation was fitted with several Lorentzians in the "$\nu_{max}$" representation \citep*{belloni2002}. The fit parameters were: the characteristic frequency $\nu _{\rm max} \equiv \nu _{0}\sqrt{1+1/(4Q^{2})}$, the quality factor Q $\equiv\nu_{0}$/FWHM, and the integrated power $P$, where $\nu_{0}$ is the centroid frequency and FWHM is the full width at half maximum of the Lorentzian. When Q turned out negative, it was fixed to 0 (i.e., we fitted a zero-centred Lorentzian); this did not significantly affect the other parameters. We only report components with a single-trial significance $P/\sigma_{P_-}$ $>$ 3.0,  with $\sigma_{P_-}$ the negative error on $P$ calculated using $\Delta\chi^2$ = 1. 

\section{Results}\label{results}
\subsection{Light curves and color diagram}\label{lchd}
\noindent Figure ~\ref{lchc}  shows the light curve and hardness-intensity diagram (HID) of the 2005-10 observations. Intensity is the count rate in the respective energy band, while hardness (or ``hard color'') is the 2--10/0.5--1 keV intensity ratio.  The 2005 light curve was fast-rise exponential-decay, consistent with the RXTE results (C07, Z07, RS07 and Ch10). The intensity peaked at 152~c/s and 59~c/s in the soft and hard band, respectively. By MJD~53605 it had decayed to below 20~c/s per band, where it remained throughout 2006-08. From the peak of the outburst in 2005 until 2007, the source gradually became harder, as can be seen in the HID, with the first observation in 2007 (MJD~54248.7) the hardest; thereafter it gradually softened. When the source was observed again in 2010,  the intensity had increased to 100~c/s in the soft band while  in the hard band it was below 20~c/s. So, consistent with the report by \citealt{soleri2012}, the source was much softer, as can also be seen in the HID (hardness $<$ 1), with the observation on MJD~55337.2 the softest. We occasionally refer to the spectral state in 2010 as soft without implying any particular canonical state viz. High Soft State or Very High State.

\begin{figure}
\center
\includegraphics[width=7cm,height=10cm,angle=-90]{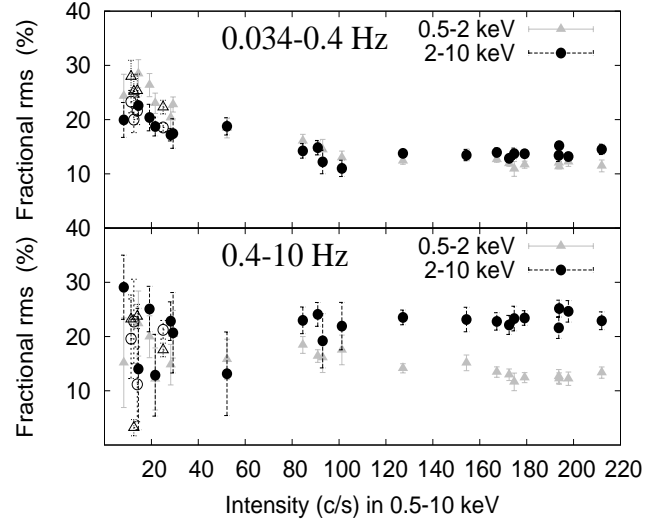}
\caption{Dependence of fractional rms amplitude on intensity in energy bands and frequency ranges as indicated. The filled points represent the individual observations in 2005 and the empty points represent the averages over MJD 53638-53640 (in 2005) and of  2006, 2007 and 2008 in the respective energy bands. }\label{rms}
\end{figure}
\begin{figure}
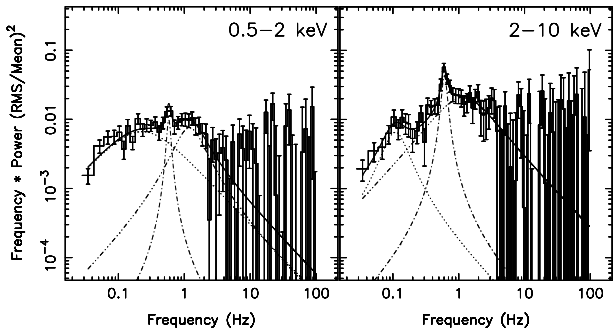

\centering
\includegraphics[width=4.425cm]{0.5-2pds.ps}\includegraphics[width=3.65cm]{2-10pds.ps}
\caption{The rms normalized power spectra of an XRT observation (00030090015, MJD~53562.35) in the peak of the outburst, in soft (0.5--2 keV, 90~c/s) and hard (2--10 keV, 37~c/s) bands. The best fit model using three Lorentzians is shown in each frame.}\label{2005pds}
\end{figure}
\subsection{Timing analysis}\label{timeanalysis}
\subsubsection{Model-independent analysis}
\noindent We first investigate the energy dependence of the variability in a model-independent way. We calculated the fractional rms amplitude from the power spectra by integrating the power over the frequency ranges 0.035--0.4~Hz and 0.4--10~Hz (chosen based on the frequencies of the power spectral components, see below) separately in the hard and soft band. Suppression of the fractional rms in the CCD data affected by pile-up has been reported earlier with \textit{Chandra} data \citep{tomsick2004}. We observe similar changes of up to 1.1\% rms (see the Appendix for details). The amplitudes obtained are consistent within errors before and after removal of the piled-up data and do not affect our conclusions. However, we remove the piled-up data from our analysis as the energy of the piled-up photons cannot be estimated and this information is necessary for our energy dependent study. Figure ~\ref{rms} shows the rms over these two frequency ranges and energy bands as a function of intensity (corrected for all artefacts) for individual observations up to MJD~53638, and for the average power spectra of  observations in MJD 53638-53640 and the years 2006, 2007 and 2008 (as the rms is not well constrained in individual observations in the frequency ranges). The 2010 data has very little variability (as expected in a soft state) which is poorly constrained and hence not plotted. \\ 
\\
At low intensity ($<$40~c/s), the $<$0.4 Hz variability is a bit weaker in the hard band (18--24\% rms) than in the soft band (20--30\%), whereas the $>$0.4~Hz variability has large error bars but is consistent with being equal in the two bands at 10--30\%.  This is consistent with what \cite{wilkinson} found in the 2006 \textit{XMM-Newton} data.  \\
\\
The \textit{XMM-Newton} observation was taken when the source had decayed to low intensity. \textit{Swift} also covered the peak of the outburst. Hence, we now report on what happened in the peak of the outburst, i.e., at high intensity ($>$40~c/s): the $<$0.4~Hz variability in the hard band is systematically slightly higher than in the soft band for intensity  $>$120~c/s (it is similar in both bands between 40 and 120 c/s). Going from low to high intensity, the behavior swaps between the two energy bands. The $>$0.4~Hz variability above 80 c/s is clearly {\it stronger} in the hard band (20--25\%) than in the soft band (12-18\%). Both $<$0.4~Hz and (using also the information from \textit{XMM-Newton}, \citealt{wilkinson}) $>$0.4~Hz the fractional rms spectrum (rms as a function of energy) softens as the source decays, contrary to the overall source spectrum, which gets harder. In terms of differences in fractional rms between the peak of the outburst and the decay, the $<$0.4~Hz variability increases as the source decays, but more strongly in the soft band than in the hard band. The $>$0.4~Hz variability shows little evidence of change (the error bars at low intensity are large), although in the soft band there is some evidence for an increase in the rms as the source drops below 120~c/s. 
\subsubsection{Power-spectral fits}
\begin{figure}
\includegraphics[width=6cm,height=6.5cm]{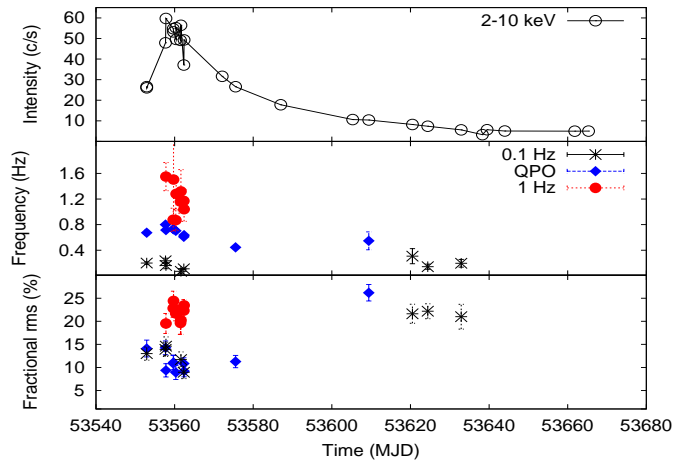}
\caption{Intensity, frequency and fractional rms amplitude of the components detected in the power spectra in the hard band as a function of time in 2005. Each point represents one observation.  Symbols as indicated.
}\label{2-10-2005}
\end{figure}
\begin{figure}
\center
\includegraphics[width=6cm,height=6.5cm]{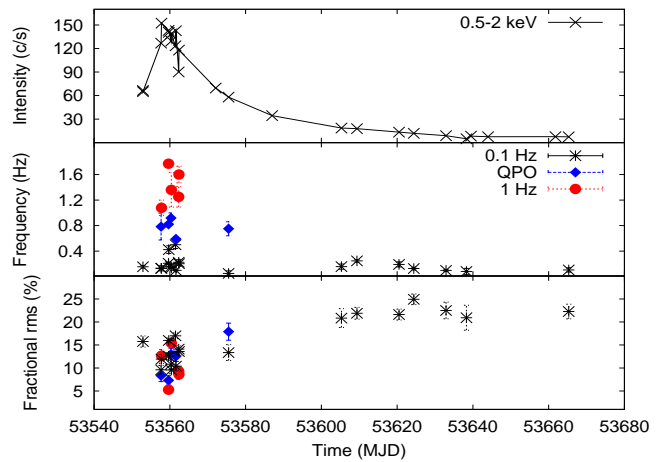}
\caption{Intensity, frequency and fractional rms amplitude of the components detected in the power spectra in soft band as a function of time in 2005. Each point represents one observation.  Symbols as indicated.}\label{0.5-2-2005}
\end{figure}
\noindent We now investigate how the energy dependence of the variability, and in particular the extra variability $>$0.4~Hz we found in the hard band during the peak of the outburst, translates into energy dependencies of the power-spectral components. We analyse the power spectra of all observations in the hard and soft band. Figure ~\ref{2005pds}, shows  representative power spectra of an observation in the peak of the outburst.  Three components are detected, which in  individual observations during 2005 vary in characteristic frequency over the ranges 0.07--0.3~Hz (at Q=0--2.1), 0.5--0.9~Hz (Q=1--11) and 0.8--1.6~Hz (Q=0--0.8).  Henceforth, we refer to them as the 0.1~Hz component, the QPO and the 1~Hz component, respectively.  Figure ~\ref{2-10-2005} shows the light curve, and the evolution of frequency and amplitude of the components during 2005 in the hard band.  As a sanity check we compared our \textit{Swift} XRT results in the hard band with earlier RXTE results. Similar power spectra have also been observed with RXTE (Figure 4 in Z07, RS07 and C07). As only the frequency evolution of the QPO was presented in Z07, RS07, C07 and Ch10, we cannot check the other parameters, but our QPO frequencies match well to the RXTE ones.\\ 
\\
The RXTE PCA has a larger effective area than the \textit{Swift} XRT, so not all components observable with RXTE might be detected in individual \textit{Swift} observations. To investigate this, we averaged the power spectra over MJD~53552--53580 (hereafter the ``outburst-peak'') in the 0.5--10 keV band, for a total exposure of $\sim$14 ksec.  One additional component was detected, at 1.5~Hz,  but its single-trial significance is only 2.5$\sigma$. This component is similar to one that can be seen in the power spectra plotted in Z07, RS07 and C07 (Figure 4). We conclude that our \textit{Swift} results in the 2-10 keV band are consistent with RXTE, although at reduced sensitivity, exactly as expected.  \\
\begin{figure}
\center
\includegraphics[width=6cm,height=6.5cm]{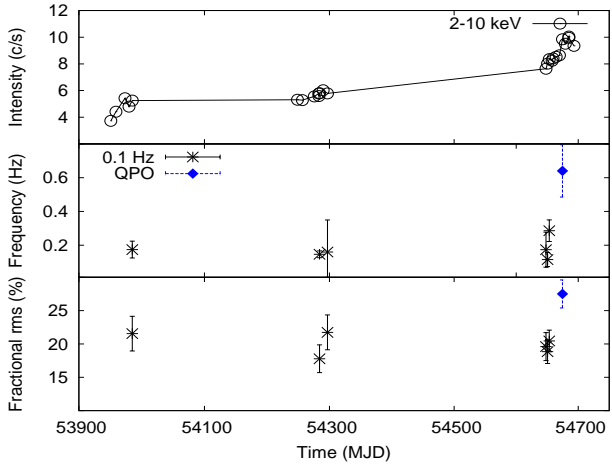}
\caption{Intensity, frequency and fractional rms amplitude of the components detected in the power spectra in the hard band as a function of time during 2006-08. Each point represents one observation.  Symbols as indicated.} 
\label{2-10-2006-09}
\end{figure}
\begin{figure}
\center
\includegraphics[width=6cm,height=6.5cm]{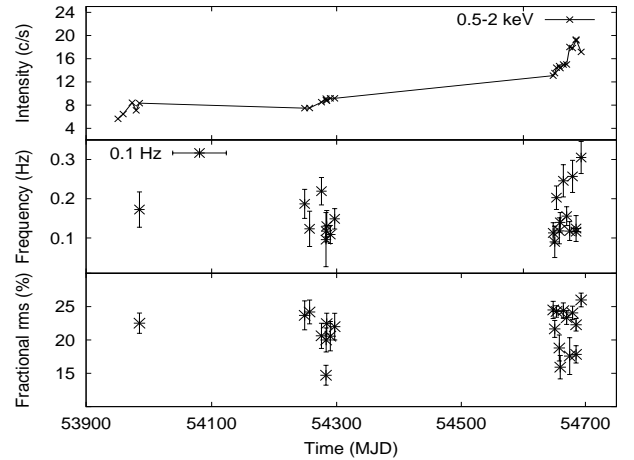}
\caption{Intensity, frequency and fractional rms amplitude of the components detected in the power spectra in the soft band as a function of time during 2006-08. Each point represents one observation. Symbols as indicated.} 
\label{0.5-2-2006-09}
\end{figure}

\begin{figure}
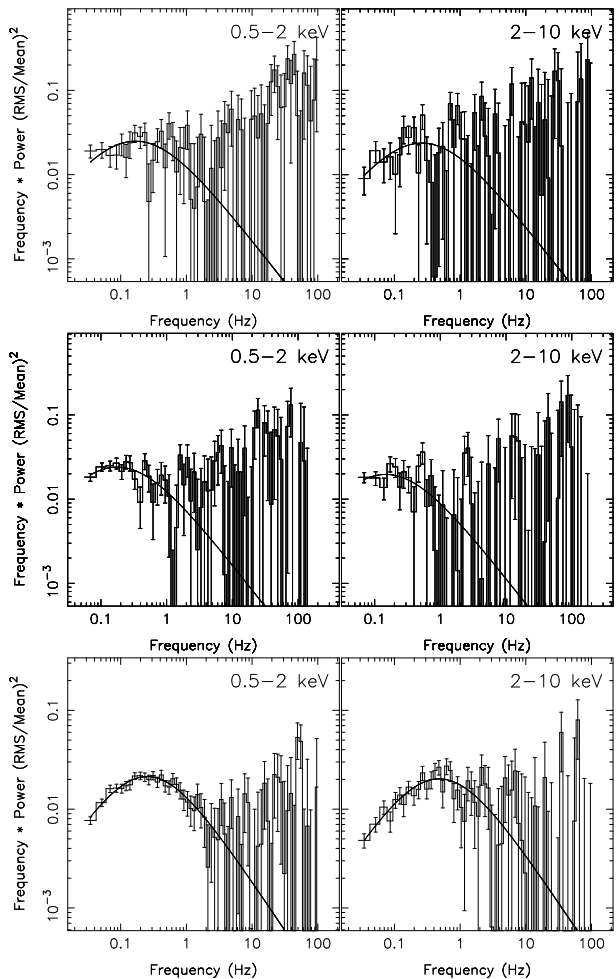

\includegraphics[width=4.41cm]{0.5-2-2006pds.ps}\includegraphics[width=3.64cm]{2-10-2006pds.ps}
\includegraphics[width=4.41cm]{0.5-2-2007pds.ps}\includegraphics[width=3.64cm]{2-10-2007pds.ps}
\includegraphics[width=4.41cm]{0.5-2-2008pds.ps}\includegraphics[width=3.64cm]{2-10-2008pds.ps}
\caption{The rms normalized average power spectra for the years (top to bottom) 2006, 2007 and 2008 in hard and soft bands as indicated. The best fit model using one Lorentzian is shown in each frame. The power at high frequencies seen in some of the spectra above is not significant.} \label{pds2006-09}
\end{figure}

\noindent We now look at the evolution of the power spectral components in 2005 in the soft band and compare it with the hard band. A representative power spectrum is shown in Figure \ref{2005pds}, left panel. Figure \ref{0.5-2-2005} shows the light curve, evolution of frequency and the rms of these components with time. The most obvious difference between the two energy bands is that in the outburst-peak, while in the soft band all components vary over similar strength (5-18\%), in the hard band the 1~Hz component is clearly stronger (20-25\%) than the other components.\\
\\
Figures ~\ref{2-10-2006-09} and \ref{0.5-2-2006-09} show the evolution for 2006-08 in the hard and soft band, respectively. The 0.1~Hz component is present, at similar frequency and amplitude in both bands, but there is only a single detection of the QPO (in the hard band), and the 1~Hz component is not detected. The 2006, 2007 and 2008 average power spectra of both bands are shown in Figure ~\ref{pds2006-09}. In 2010 there is no significant variability in either band ($<$ 3 \%). This is consistent with the source being in a softer, less variable state.   \\

\begin{figure}
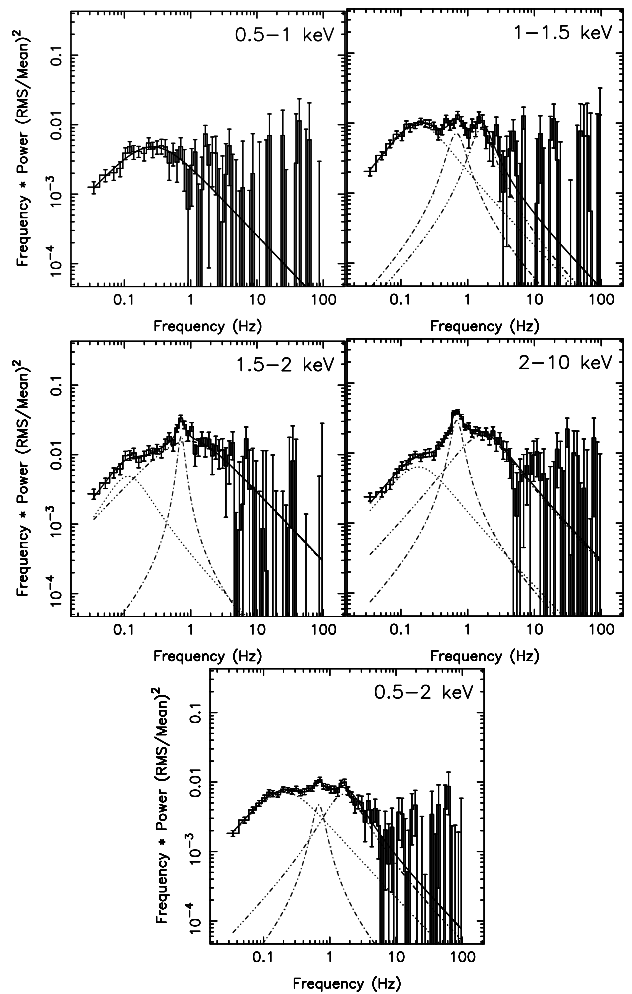

\center
\includegraphics[width=4.45cm]{0.5-1-2005pds.ps}\includegraphics[width=3.7cm]{1-1.5-2005pds.ps}
\includegraphics[width=4.45cm]{1.5-2-2005pds.ps}\includegraphics[width=3.7cm]{2-10-2005pds.ps}
\includegraphics[width=4.45cm]{0.5-2-2005wypds.ps}
\caption{The rms normalized average power spectra of the outburst-peak (MJD~53552--53580) in energy bands as indicated. The best fit model using three Lorentzians are shown. Corresponding parameters are presented  in Table \ref{tab:combob}.}\label{pds2005}
\end{figure}

\noindent Earlier studies  \citep[e.g.,][in the BHB XTE J1550--564]{homan1550} have shown that the QPOs are stronger at higher energies. We investigate if the different components in the power spectrum have the same energy dependence based on the outburst-peak data. First we calculated power spectra separately in the hard and soft band and fitted these in two ways: with the $\nu_{max}$ and Q of the respective components ``tied'' (constrained to be the same) and only their integrated powers free, and with all the parameters free. Both fits have a reduced $\chi^2$ of $\sim$1.03 (with 11560 and 11554 degrees of freedom, respectively),  and similar values within errors for tied and untied parameters, except for the Q of the 1~Hz component, which in the tied fit is 0.80$\pm$0.08, and in the untied fit 1.19$\pm$0.15 in the soft, and 0.66$\pm$0.12 in the hard band. The rms values of the QPO and the 1 Hz component are higher in the hard band than the soft band (both by $\sim$9 \%) in both tied and untied fits. This shows that the high frequency components are stronger at higher energies. The harder 1 Hz component and QPO  cause the extra power $>$0.4 Hz reported above in Figure \ref{rms}. The 1 Hz component is also broader in the hard band and perhaps should be fitted independently. Hence, we fitted the power spectra of all individual observations (discussed above) independently in both bands with all parameters free.\\
\\
Finally, to study the energy dependencies $<$2~keV, we calculated the outburst-peak power spectra in the 0.5--1 keV, 1--1.5 keV and 1.5--2 keV bands.  All the outburst-peak power spectra are collected  in Figure ~\ref{pds2005} and their best-fit parameters in Table~\ref{tab:combob}. The 0.1~Hz component is detected with similar strength (11-15\% rms), frequency and Q in all bands. The QPO is not constrained in the 0.5--1 keV band, it is $\sim$8\% rms in both the 1--1.5 keV and 1.5--2 keV bands, and nearly twice that amplitude again in the 2--10 keV band, indicating that this component is hard. The 1~Hz component is  not constrained in the 0.5--1 keV  band, is detected at $\sim$9\% in the 1--1.5 keV band and more than doubles in amplitude to $\sim$20\% rms in the 1.5--2 keV and 2--10 keV bands. The low frequency component is strong in all bands, while the high frequency components get stronger as we go to higher energies. \\

\subsection{Correlations}\label{correlations}
\begin{figure}
\center
\includegraphics[width=5cm,height=8.5cm,angle=-90]{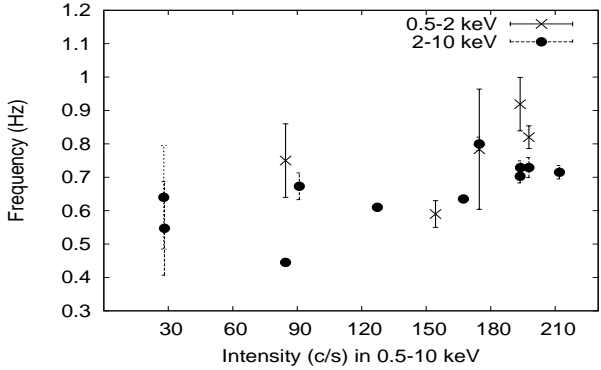}
\caption{The correlation of QPO frequency with intensity during 2005-08. The energy bands are indicated.}\label{freq-in-0.8hz}
\end{figure}

\noindent We looked for correlations between power spectral parameters and intensity. The hardness is inversely correlated with intensity (see Figure \ref{lchc}) and correlations of hardness with  power spectral parameters does not give additional information. Hence, we show the correlations with intensity only. We first inspect the dependence of the {\it characteristic frequency} of the components on intensity.  The 0.1~Hz component shows no correlation with intensity in either energy band at any time. As seen in Figures. \ref{2-10-2005}, \ref{0.5-2-2005},  \ref{2-10-2006-09} and \ref{0.5-2-2006-09} (stars) the component is present during 2005-08 over a wide range of intensities, but always at similar frequencies.   The QPO frequency as a function of intensity is plotted in Figure ~\ref{freq-in-0.8hz}. In the hard band (circles) there is some evidence of a positive  correlation, as reported earlier by Z07 and RS07.  In the soft band (crosses) there are few detections, and the dependence of frequency on intensity is not clear. The 1~Hz component also has very few detections, and no clear correlation is seen. For each component, Q covered a wide range of values, as reported in Section ~\ref{timeanalysis}. We inspected the dependence of Q on frequency, intensity and rms, but there were no clear correlations.\\
\\

\begin{figure}
\center
\includegraphics[width=5cm,height=8.5cm,angle=-90]{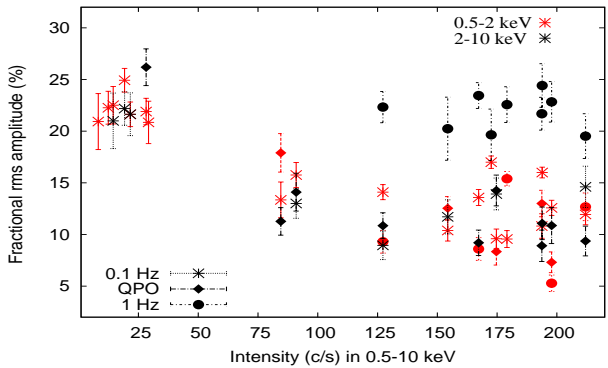}
\caption{The fractional rms amplitudes of the power spectral components as a function of intensity during 2005-08. The energy bands and symbols are as indicated.}\label{rms-in}
\end{figure}

\noindent The rms of the individual components for 2005-08 is shown in Figure ~\ref{rms-in} as a function of intensity. For both the 0.1~Hz component (stars) and the QPO (diamonds) and in both energy bands, rms is anti-correlated to intensity. As there are no detections of the 1~Hz component (circles) at low intensities, its rms-intensity correlation is not clear. The 1~Hz component stands out in the hard band as the strongest of all components in the peak. 
\begin{table}
\center
{\renewcommand{\arraystretch}{}
\begin{tabular}{|c|c|c|c|c|}
\hline 
Energy & $\nu_{max}$ & Q & rms (\%)& rms (\%) \\
(keV) & Hz& & & 0.03-10~Hz\\
\hline
& 0.23 $\pm$ 0.07 & 0.08 $\pm$ 0.20 & 11.9 $\pm$ 1.4 &\\
0.5--1 & -- & -- & -- & 13.0 $\pm$ 1.1  \\
& -- & -- & -- &\\
\hline
& 0.18 $\pm$ 0.02 & 0.30 $\pm$ 0.10 & 15.1 $\pm$ 1.0 &\\
1--1.5 & 0.67 $\pm$ 0.05 & 1.34 $\pm$ 0.79 & 8.8 $\pm$ 2.1 & 19.3 $\pm$ 1.3\\
& 1.56 $\pm$ 0.10 & 1.60 $\pm$ 0.72 &  8.7 $\pm$ 1.3 &\\
\hline
& 0.12 $\pm$ 0.03 & 0.45 $\pm$ 0.32 & 10.7 $\pm$ 2.2 &\\
1.5--2 & 0.72 $\pm$ 0.02 & 3.91 $\pm$ 1.83 & 8.9 $\pm$ 2.3 & 27.1 $\pm$ 1.3\\
 & 1.07 $\pm$ 0.34 & 0.0 (fixed) & 22.8 $\pm$ 1.5 &\\
 \hline
& 0.18 $\pm$ 0.03 & 0.23 $\pm$ 0.15 & 13.0 $\pm$ 1.3 &\\
2--10 & 0.69 $\pm$ 0.01 & 1.97 $\pm$ 0.36 & 15.1 $\pm$ 1.5 & 27.2 $\pm$ 1.1\\
& 1.79 $\pm$0.23 &  0.45 $\pm$ 0.20 & 19.1 $\pm$ 2.4 &\\
\hline
& 0.21 $\pm$ 0.02 & 0.08 $\pm$ 0.07 & 14.3 $\pm$ 0.5 &\\
 0.5--2 & 0.69 $\pm$ 0.03 & 1.80 $\pm$ 0.70 & 6.3 $\pm$ 1.0 & 18.7 $\pm$ 1.0\\
 & 1.6 $\pm$ 0.6 & 0.91 $\pm$ 0.25 & 9.5 $\pm$ 0.8 &\\
\hline
 \end{tabular}}
\caption{Best fit parameters of the averaged power spectra in the peak of the outburst in different energy bands shown in Figure ~\ref{pds2005}. The rms (\%) reported are the fractional rms amplitudes of: 4th column - the individual components, and, 5th column- integrated over 0.03-10 Hz.}\label{tab:combob}
\end{table}
 \\
 
\section{Discussion}\label{discussion}
\noindent We report on the energy dependent variability of the first six years of the very long and still ongoing outburst of the BHB SWIFT J1753.5-0127. We observe the Fast Rise Exponential Decay type light curve also reported in earlier works (C07, RS07, Z07 and Ch10). The HID does not resemble in either \textit{Swift} or RXTE (\citealt{soleri2012}) data to the typical "q"-shaped track followed by many other BHBs. The source remained in hard states throughout the outburst (C07, RS07, Z07, Ch10), until a transition to a softer state occurred in 2009 (\citealt{soleri2012}), leading to our softest observations in 2010. As we performed very detailed power spectral studies with the \textit{Swift} XRT, we first examined the behavior of the Poisson level and find it to be affected by instrumental effects (see the Appendix for details). We carefully  compared our results with the earlier RXTE results using the 2--10 keV band, and found them to be consistent.  \\
\\
The "type-C" QPO (peaked low frequency components accompanied by strong broadband noise in the intermediate states, e.g. \citealt{casella2005}), the  break frequency, $\nu_b$ and the hump frequency, $\nu_h$, reported by \citealt{soleri2012} correspond to the QPO, the 0.1 Hz and 1 Hz components, respectively, in our analyses. We report  these components,  for the first time, in soft X-rays together with the harder band. The WK correlation, which is a positive correlation between  $\nu_b$  and $\nu_h$  \citep{wk} was reported by \citealt{soleri2012} between these components in SWIFT J1753.5-0127.  As there are very few simultaneous detections of these components in our data, we do not see a clear WK correlation in either band  and is hence not shown here. The fractional amplitudes of all three components vary over similar range (5-18 \%) in both bands, except the 1 Hz component in the hard band which is the strongest (20-25 \%) amongst all components in the peak of the outburst. In the decay, all the components get stronger in both energy bands, with no detections of the 1 Hz component. The analysis in soft sub-bands and hard band (see Figure \ref{pds2005} and Table \ref{tab:combob}) in the peak of the outburst shows that the 0.1 Hz component displays a flat rms spectrum i.e. the fractional amplitudes do not vary with energy. The QPO and the 1 Hz component cannot be constrained in the 0.5--1 keV band. These get stronger at higher energies, with the QPO getting stronger above 2 keV while the 1 Hz component gets stronger above 1.5 keV.    \\
\\
The main new finding of our work (Figure \ref{rms}) is that in the peak of the outburst (for $>$120~c/s) the variability is weaker in the soft band  than in the hard band; the difference is most pronounced at high ($>$0.4~Hz) frequency but also seen at low ($<$0.4~Hz) frequency. Previous work with \textit{XMM-Newton} \citep{wilkinson} had shown that in the decay, in 2006, the opposite was true: at low frequency the soft band was more variable than the hard band, while at higher frequencies both bands had similar fractional variability amplitudes. Our 2005-2008 results in the decay are consistent with this.  \citet{wilkinson} interpreted this extra soft-band variability in terms of a stronger contribution of the soft disk emission to the variability at low frequencies, which could be associated with accretion fluctuations intrinsic to the disk at these frequencies.  Combining our new results on the outburst-peak with these earlier findings on the decay, we can say that in both frequency ranges the fractional rms spectrum becomes softer as the source decays, while instead the overall source spectrum becomes harder.\\  
\\
These findings can be understood in terms of a simple two-component representation (hot inner flow and cool disk) of the picture of \citet{lyub1997} and \citet{chu2001}.  The softer overall spectrum in the peak of the outburst is due to a relatively stronger disk contribution. That the fractional rms spectrum gets harder at all frequencies when the disk contribution increases can be explained if the observed disk variability has a lower fractional rms than the hot flow: the increased disk flux dilutes the variability at low energy.  This would imply that while in the decay the disk at low frequencies  is more variable (in fractional rms) than the hot flow, in the outburst-peak it is {\it less} variable.  So, we may be seeing in this outburst-peak a transition to the less variable disk state that must also characterize the soft states. That in our outburst-peak data the difference between low and high energy variability is less at low frequency can be due to a residual contribution of intrinsic disk variability at low frequency similar to that proposed by \citet{wilkinson} and \citet{uttley2011}. \\
\\
It would be interesting to observe the variability behavior of this source when (if) it makes a transition back to the hard states, to see if it exhibits behavior similar to other BHBs during this transition, even if up to now it did not follow the usual "q" path in the HID. With our results, we demonstrate that the power spectral studies can be successfully performed with \textit{Swift} XRT. The soft band provides very useful insight into the energy dependence of the variability which is essential to understand the contribution of the accretion disk. To understand better the role of the disk, variability studies of more BHBs with \textit{Swift} XRT are required. We recommend caution while dealing with the  Poisson spectrum of the power spectra generated with \textit{Swift} XRT. 

\section*{Acknowledgements}
We thank the anonymous referee for his/her useful comments that greatly helped  to improve the manuscript.We thank Kim Page for her help in \textit{Swift} XRT data reduction issues. We would like to thank P Cassatella for his helpful discussion. This research has made use of data obtained from the High Energy Astrophysics Science Archive Research Center (HEASARC), provided by NASA's Goddard Space Flight Center, and also made use of NASA's Astrophysics Data System. \\
\appendix{
{The XRT is a grazing incidence Wolter 1 telescope which focuses the incoming 0.3--10 keV X-rays onto a CCD of 600$\times$602 pixels (23.6 $\times$23.68 sq.arc-min). For a good time resolution (1.766 ms) and to mitigate pile-up in bright sources, the CCD is operated in the WT mode.  In this mode, in wt2 configuration, only the central 200 columns of the CCD are read out. 10 pixels are binned along columns  and hence the spatial information is lost in this dimension \citep{burrows2005}.\\
\\
\begin{figure}
\center
\includegraphics[width=6.5cm,angle =-90]{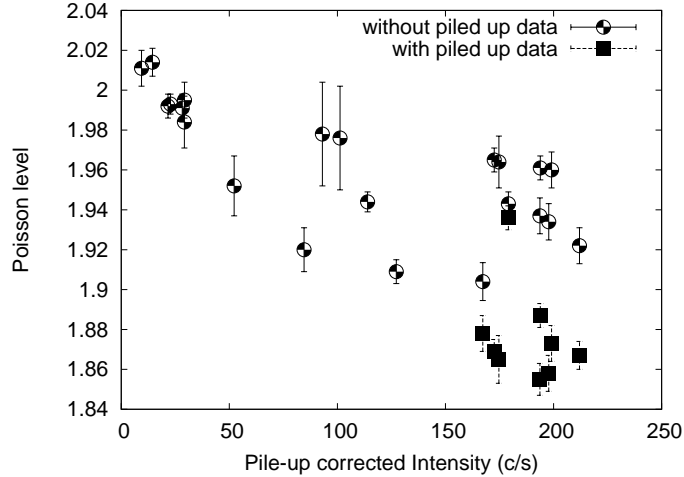}
\caption{ The Poisson noise level measured between 50--100~Hz in Leahy-normalized power spectra in the 0.5-10 keV band as a function of intensity. The intensity on X-axis is corrected for pile-up and bad column and background subtracted. As pile-up affects the data above 150 c/s, the circles above 150 c/s are observations where the Poisson level is measured in the power spectra generated after removing the piled up data, while the filled squares are same observations where the Poisson level is measured in the power spectra generated without the removal of the piled up data. It can be seen that the Poisson level improves after the removal of the piled up data.}\label{poisson}
\end{figure}
\begin{figure}
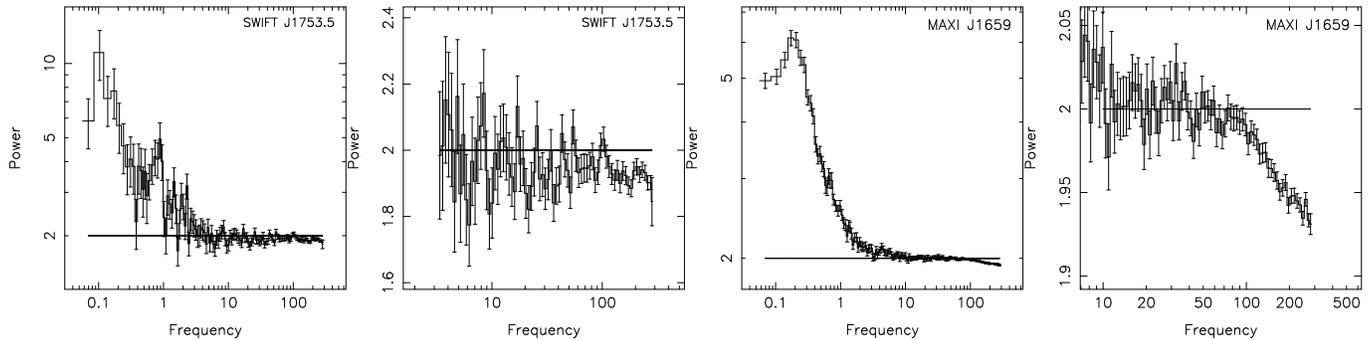

\includegraphics[width=4.5cm]{1753pds.ps}\includegraphics[width=4.5cm]{1753pds-z.ps}\includegraphics[width=4.5cm]{pds-maxi.ps}\includegraphics[width=4.5cm]{pds-maxiz.ps}
\caption{Leahy normalized power spectra of two sources (as indicated) in the 0.5--10 keV band over the full frequency range and at higher frequency. Horizontal lines at power level 2.0 indicate the expected Poisson level in the absence of instrumental effects.}\label{pdsapp}
\end{figure}
\noindent 
The arrival of photons follows a Poisson distribution. In a Leahy normalized power spectrum, we expect the Poisson noise level at  2.0 \citep{leahy1983}. However, in the case of pile-up the events are no longer independent, so this does not hold true any more.  Pile-up occurs when multiple X-ray photons incident on a  3 $\times$ 3 pixel region are read out as a single event of higher energy \citep{romano2006}. The effects of pile-up on the Poisson noise spectrum have not been systematically studied yet for \textit{Swift} XRT CCD in the WT mode. We examined the effects on the Poisson noise level of pile-up, which  for the XRT CCD is expected to appreciably affect the data above 150 counts/sec.\\ 
\\
Power spectra were obtained using the same method as discussed in Section 2. We used observations of SWIFT J1753.5-0127 covering count rates up to 210~c/s. Over the frequency range 50--100~Hz, where no variability is observed with \textit{Swift}, we fit a constant to measure the Poisson level.  Figure. \ref{poisson} shows the Poisson level as a function of count rate. It can be clearly seen that Poisson level decreases as the count rate increases. For the observations with count rates above 150~c/s, when the piled-up data is included in the power spectra, the Poisson level goes down to 1.86. If we remove the piled-up data and then generate the power spectrum in the same observations, the Poisson level improves and is above 1.90. However, it is worth noting that the Poisson level is already less than 2.0 below the nominal pile-up threshold of 150 c/s. So, even at much lower count rates, fixing the Poisson level at 2.0 is not good practice. This should be taken into account when fitting \textit{Swift} XRT power spectra.  We also investigated if there was any dependence of the Poisson level on photon energy, but no obvious dependence was found. \\
\\
Suppression of the fractional rms amplitude in the data affected by pile up has been reported earlier in \textit{Chandra}  CCD data \citep{tomsick2004}. The amplitude drop due to pile-up observed by them was about 1\%. They also report the effects of pile-up on the Poisson level and that the pile-up does not affect the shape of the power spectrum, consistent with what we observe. To inspect the effects on the rms amplitudes, we calculate the amplitudes in  the 0.5-10 keV energy band up to 10 Hz in the observations with intensity above 150 c/s. We calculate this for two power spectra per observation: one generated including the data affected by pile-up and one excluding the pile-up affected data (same method discussed in section \ref{obdata}). We also observe a difference in the amplitudes of up to 1.1 \% rms.\\
\\
\noindent While inspecting the power spectra we also observed a power drop-off at frequencies above 100~Hz. Figure ~\ref{pdsapp} shows the power spectra of  observations of the sources SWIFT J1753.5-0127 and MAXI J1659.5-152 (another BHB) over the full frequency range and at higher frequency in the 0.5-10 keV band. The constant line at value 2 shows the expected Poisson level in an ideal case. A significant power drop-off is seen above 100 Hz, more clearly in the case of MAXI J1659.5-152. The drop-off is energy dependent and becomes stronger as the energy increases. This can be explained in terms of the size of the event formed by the interaction of the X-ray photon on the CCD pixels. Soft photons form predominantly single pixel events, while the hard photons form multiple pixel events and are hence susceptible to splitting at the 10 row boundaries during the read out in WT mode\footnote{See $http://www.swift.ac.uk/analysis/xrt/digest\_cal.php\#pow$ for details. It has also been noted that this readout method causes a slight increase in the noise level below 100 Hz, which can be seen in the simulated WT power spectra on this webpage.}. As it is not possible to identify and eliminate these events, in our analysis we put a cut-off at 100~Hz for all power spectra. \\
\\}

\begin{thebibliography}{31}
\expandafter\ifx\csname natexlab\endcsname\relax\def\natexlab#1{#1}\fi
\expandafter\ifx\csname url\endcsname\relax
  \def\url#1{{\tt #1}}\fi
\expandafter\ifx\csname urlprefix\endcsname\relax\def\urlprefix{URL }\fi

\bibitem[{{Belloni} et~al.(2002){Belloni}, {Psaltis}, \& {van der
  Klis}}]{belloni2002}
{Belloni} T., {Psaltis} D., {van der Klis} M., 2002, \apj, 572, 392

\bibitem[{{Belloni} et~al.(2005){Belloni}, {Homan}, {Casella}
  et~al.}]{belloni2005}
{Belloni} T., {Homan} J., {Casella} P., et~al., 2005, \aap, 440, 207

\bibitem[{{Burrows} et~al.(2005){Burrows}, {Hill}, {Nousek}
  et~al.}]{burrows2005}
{Burrows} D.N., {Hill} J.E., {Nousek} J.A., et~al.,  2005, \ssr, 120, 165

\bibitem[{{Cadolle Bel} et~al.(2007){Cadolle Bel}, {Rib{\'o}}, {Rodriguez}
  et~al.}]{cadolle}
{Cadolle Bel} M., {Rib{\'o}} M., {Rodriguez} J., et~al., 2007, \apj, 659,
  549

\bibitem[{{Casella} et~al.(2005){Casella}, {Belloni}, \&
  {Stella}}]{casella2005}
{Casella} P., {Belloni} T., {Stella} L.,  2005, \apj, 629, 403

\bibitem[{{Chiang} et~al.(2010){Chiang}, {Done}, {Still}, \&
  {Godet}}]{chiang2010}
{Chiang} C.Y., {Done} C., {Still} M., {Godet} O.,  2010, \mnras, 403, 1102

\bibitem[{{Churazov} et~al.(2001){Churazov}, {Gilfanov}, \&
  {Revnivtsev}}]{chu2001}
{Churazov} E., {Gilfanov} M., {Revnivtsev} M.,  2001, \mnras, 321, 759

\bibitem[{{Curran} et~al.(2011){Curran}, {Maccarone}, {Casella}
  et~al.}]{curran1752-swift}
{Curran} P.A., {Maccarone} T.J., {Casella} P., et~al., 2011, \mnras, 410,
  541

\bibitem[{{Done} et~al.(2007){Done}, {Gierli{\'n}ski}, \& {Kubota}}]{done2007}
{Done} C., {Gierli{\'n}ski} M., {Kubota} A.,  2007, \aapr, 15, 1

\bibitem[{{Evans} et~al.(2007){Evans}, {Beardmore}, {Page} et~al.}]{evans2007}
{Evans} P.A., {Beardmore} A.P., {Page} K.L., et~al.,  2007, \aap, 469, 379

\bibitem[{{Hiemstra} et~al.(2009){Hiemstra}, {Soleri}, {M{\'e}ndez}
  et~al.}]{hiemstra2009}
{Hiemstra} B., {Soleri} P., {M{\'e}ndez} M., et~al.,  2009, \mnras, 394,
  2080

\bibitem[{{Homan} et~al.(2001){Homan}, {Wijnands}, {van der Klis}
  et~al.}]{homan1550}
{Homan} J., {Wijnands} R., {van der Klis} M., et~al., 2001, \apjs, 132,
  377

\bibitem[{{Kennea} et~al.(2011){Kennea}, {Romano}, {Mangano}
  et~al.}]{kennea1659}
{Kennea} J.A., {Romano} P., {Mangano} V., et~al.,  2011, \apj, 736, 22

\bibitem[{{Leahy} et~al.(1983){Leahy}, {Darbro}, {Elsner} et~al.}]{leahy1983}
{Leahy} D.A., {Darbro} W., {Elsner} R.F., et~al.,  1983, \apj, 266, 160

\bibitem[{{Lyubarskii}(1997)}]{lyub1997}
{Lyubarskii} Y.E.,  1997, \mnras, 292, 679

\bibitem[{{Miller} et~al.(2006){Miller}, {Homan}, \& {Miniutti}}]{miller2006}
{Miller} J.M., {Homan} J., {Miniutti} G., 2006, \apjl, 652, L113

\bibitem[{{Palmer} et~al.(2005){Palmer}, {Barthelmey}, {Cummings}
  et~al.}]{palmer2005}
{Palmer} D.M., {Barthelmey} S.D., {Cummings} J.R., et~al.,  2005, The
  Astronomer's Telegram, 546, 1

\bibitem[{{Ramadevi} \& {Seetha}(2007)}]{ramadevi}
{Ramadevi} M.C., {Seetha} S.,  2007, \mnras, 378, 182

\bibitem[{{Reis} et~al.(2010){Reis}, {Fabian}, \& {Miller}}]{reis2010}
{Reis} R.C., {Fabian} A.C., {Miller} J.M.,  2010, \mnras, 402, 836

\bibitem[{{Remillard} \& {McClintock}(2006)}]{remillard2006}
{Remillard} R.A., {McClintock} J.E.,  2006, \araa, 44, 49

\bibitem[{{Romano} et~al.(2006){Romano}, {Campana}, {Chincarini}
  et~al.}]{romano2006}
{Romano} P., {Campana} S., {Chincarini} G., et~al., 2006, Nuovo Cimento B
  Serie, 121, 1067

\bibitem[{{Soleri} et~al.(2012){Soleri}, {Mu{\~n}oz-Darias}, {Motta}
  et~al.}]{soleri2012}
{Soleri} P., {Mu{\~n}oz-Darias} T., {Motta} S., et~al.,  2012, ArXiv
  e-prints

\bibitem[{{Stella} \& {Vietri}(1998)}]{stella1998}
{Stella} L., {Vietri} M.,  1998, \apjl, 492, L59

\bibitem[{{Tomsick} et~al.(2004){Tomsick}, {Kalemci}, \&
  {Kaaret}}]{tomsick2004}
{Tomsick} J.A., {Kalemci} E., {Kaaret} P.,  2004, \apj, 601, 439

\bibitem[{{Uttley} et~al.(2005){Uttley}, {McHardy}, \& {Vaughan}}]{uttley2005}
{Uttley} P., {McHardy} I.M., {Vaughan} S., 2005, \mnras, 359, 345

\bibitem[{{Uttley} et~al.(2011){Uttley}, {Wilkinson}, {Cassatella}
  et~al.}]{uttley2011}
{Uttley} P., {Wilkinson} T., {Cassatella} P., et~al.,  2011, \mnras, 414,
  L60
  
\bibitem[van der Klis(1995)]{klis1995} van der Klis, M.\ 1995, 
The Lives of the Neutron Stars, 301 

\bibitem[{{van der Klis}(2006)}]{klis2006}
{van der Klis} M., 2006, {Rapid X-ray Variability}, 39--112

\bibitem[{{Wijnands} \& {van der Klis}(1999)}]{wk}
{Wijnands} R., {van der Klis} M., 1999, \apj, 514, 939

\bibitem[{{Wilkinson} \& {Uttley}(2009)}]{wilkinson}
{Wilkinson} T., {Uttley} P., 2009, \mnras, 397, 666

\bibitem[{{Zhang} et~al.(2007){Zhang}, {Qu}, {Zhang} et~al.}]{zhang2007}
{Zhang} G.B., {Qu} J.L., {Zhang} S., et~al., 2007, \apj, 659, 1511

\end{thebibliography}

\end{document}